\documentclass[12pt]{article}
\usepackage{graphicx}

\begin{document}

\title{Market panic on different time-scales}
\author{Lisa Borland  \\  Evnine  \& Associates, Inc.\\456 Montgomery Street, 
Suite 800, San Francisco, CA 94104,  USA \footnote{lisa@evafunds.com} \\Yoan Hassid \\Department of Statistics\\ Stanford University}
\maketitle

\abstract{Cross-sectional signatures of market panic were recently discussed on daily time scales in \cite{borland}, extended here to a study of  cross-sectional properties of stocks on intra-day time scales. We confirm specific intra-day patterns of dispersion and kurtosis, and find that the correlation across stocks increases in times of panic yielding a bimodal distribution for the sum of signs of returns. We also find that there is memory in correlations, decaying as a power law with exponent 0.05. During the Flash-Crash of May 6 2010, we find a drastic increase in dispersion in conjunction with  increased correlations. However, the kurtosis decreases only slightly  in contrast to findings on daily time-scales where kurtosis drops drastically in times of panic.  Our study indicates that this difference in behavior is  result of the origin of the panic-inducing volatility shock: the more correlated across stocks the shock is, the more the kurtosis will decrease; the more idiosyncratic the shock, the lesser this effect and kurtosis is positively correlated with dispersion.  We also find that there is a leverage effect for correlations: negative returns tend to precede an increase in correlations.  A stock price feed-back model with skew in conjunction with a correlation dynamics that follows market volatility explains our observations nicely.}

\section{Introduction}
We recently proposed a model with the goal of  understanding the evolution of the cross-sectional distribution and correlation dynamics among stock returns, (adding to the existing literature  of cross-sectional studies, e.g  \cite{kaizoji}-\cite{Sornette}). Specifically, we described  and modeled the interesting interplay between dispersion, kurtosis and cross-sectional correlations which we observed on daily time-scales. In that work, we found a strong anti-correlation between cross-sectional dispersion and cross-sectional kurtosis
in financial markets \cite{borland}. In particular we discovered that, in times of market panic, dispersion increases drastically, as does volatility, while cross-sectional kurtosis drops close to zero. In addition, various measures of cross-stock correlations increases, inducing a bi-modal distribution of the cross-sectional sum of signs of stock returns, implying that in such times all stocks tend to move up or down together. 
 To describe these findings we proposed a phase-transition model, where the dynamics of correlations responded to an external volatility shock, which could be interpreted as an increase in fear and uncertainty among market participants. This model describes very well all the stylized facts observed both in normal and panic times, on daily time-scales. 

  We now extend our study to intra-day time scales, and also probe some interesting questions which were prompted by the original work. We wish to quantify any  memory in the correlations, and also to see if there are any detectable asymmetries in the dynamics of the correlations. Indeed we find that correlation tend to increase more significantly after negative returns. We also find that the degree to which the kurtosis decreases depends on whether the volatility shock to the financial system is of an exogenous (market-wide) nature, or endogenous (more stock-specific).

\begin{figure}[t]
\label{fig1}
\includegraphics[width=4.5in]{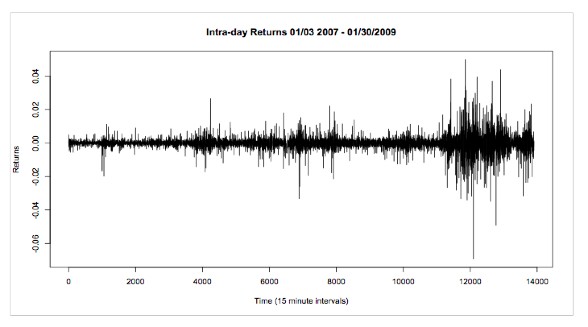} 
\caption{The market wide average of intra day returns, sampled every 15 minutes.}
\end{figure}

\begin{figure}[t]
\includegraphics[width=4.5in]{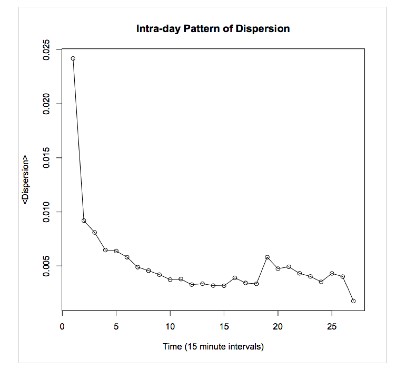} 
\caption{Average intra day dispersion}
\end{figure}

\begin{figure}[t]
\label{fig3}
\includegraphics[width=2.5in]{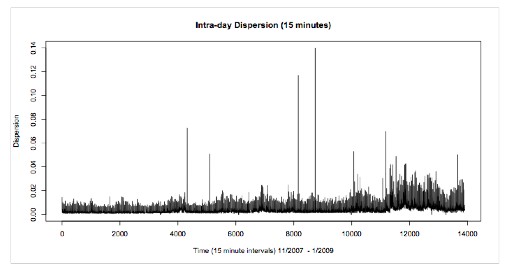} 
\includegraphics[width=2.5in]{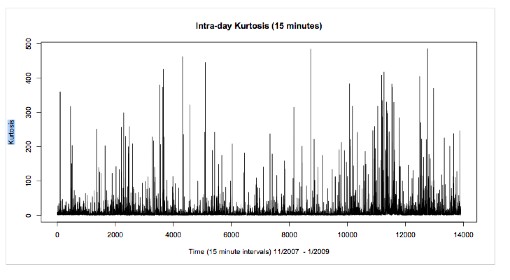} 
\caption{a) Cross-sectional dispersion calculated at 15 minute intervals and b)  cross-sectional kurtosis calculated at 15 minute intervals. In contrast to daily data (Figure 4,) \cite{borland}, there is in general no anti-correlation between kurtosis and dispersion}
\end{figure}

\begin{figure}[t]
\includegraphics[width=4.5in]{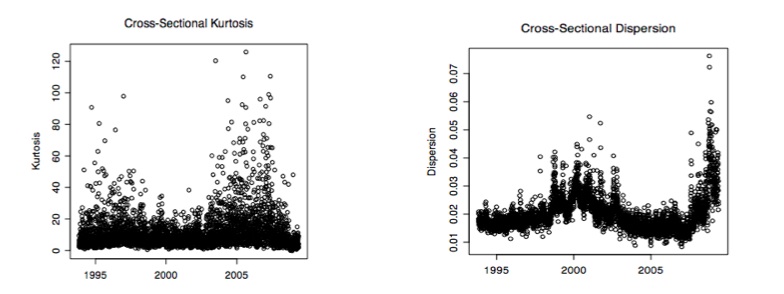}
\caption{Daily data shoes a strong anti-correlation between dispersion and kurtosis.}
\label{fig4}
\end{figure}

\begin{figure}[t]
\label{fig5}
\includegraphics[width=4.5in]{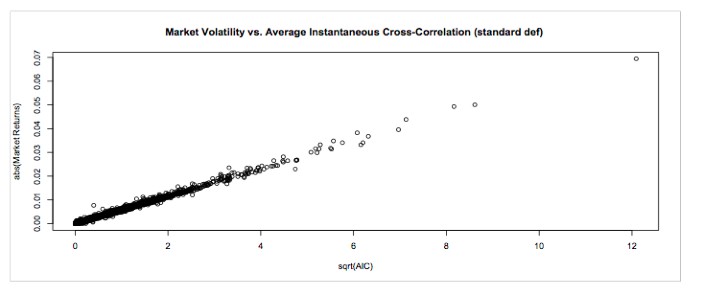} 
\caption{The definition of AIC used in \cite{multifrac} is essentially the market volatility.}
\end{figure}

\begin{figure}[t]
\label{fig6}
\includegraphics[width=4.5in]{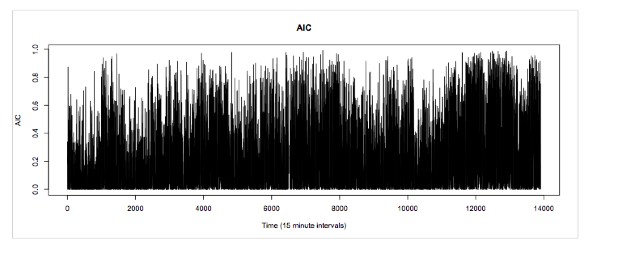} 
\caption{The cross-sectional sign correlation AIC calculated at 15 minute intervals.}
\end{figure}


\begin{figure}[t]
\label{fig7}
\includegraphics[width=4.5in]{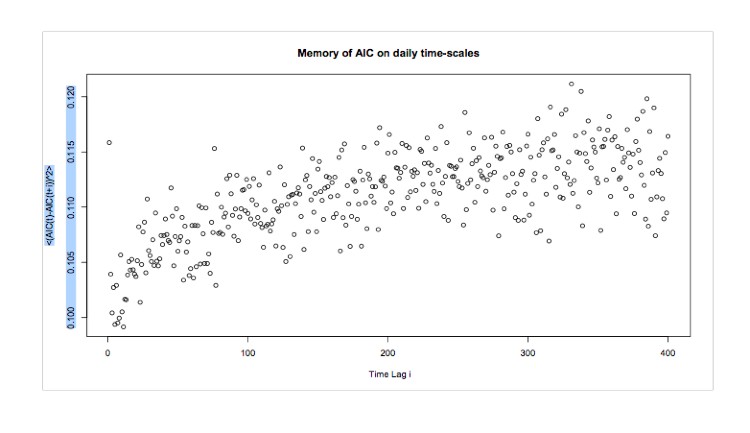} 
\caption{Memory in AIC on daily time scales.}
\end{figure}

\begin{figure}[t]
\label{fig8}
\includegraphics[width=4.5in]{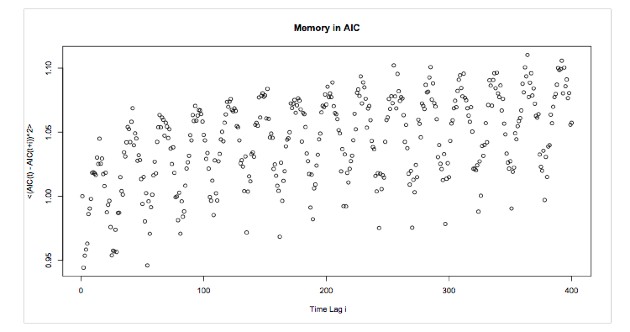} 
\caption{Memory in AIC on 15 minute time scales.}
\end{figure}

\begin{figure}[t]
\label{fig9}
\includegraphics[width=4.5in]{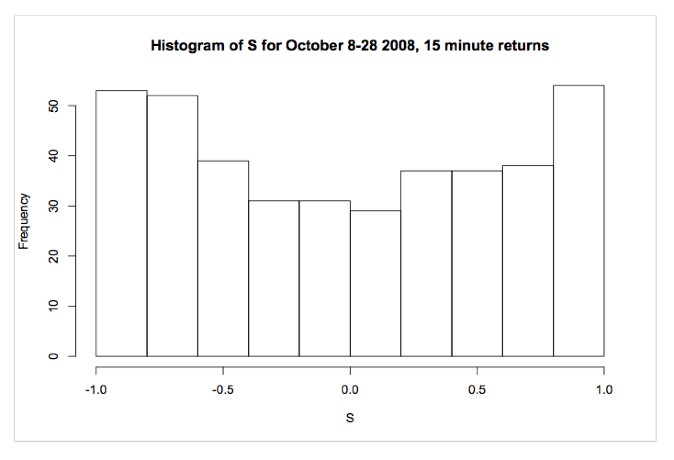} 
\caption{Histogram of sum of signs for 15 minute intervals during October 2008. Correlations are high and stocks move together.}
\end{figure}
\begin{figure}[t]
\label{fig10}
\includegraphics[width=4.5in]{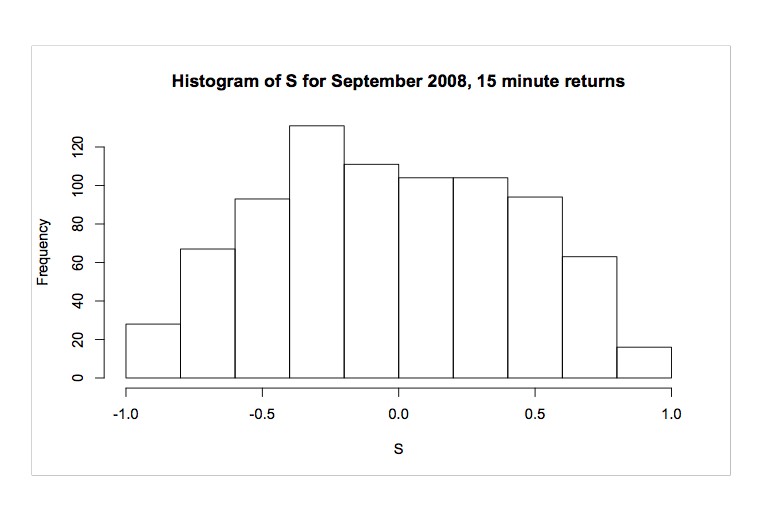} 
\caption{The distribution of the sum of signs during September 2008, taken at 15 minute intervals. The uni modal distribution implies that correlations are on average low.}
\end{figure}

\begin{figure}[t]
\label{fig11}
\includegraphics[width=4.5in]{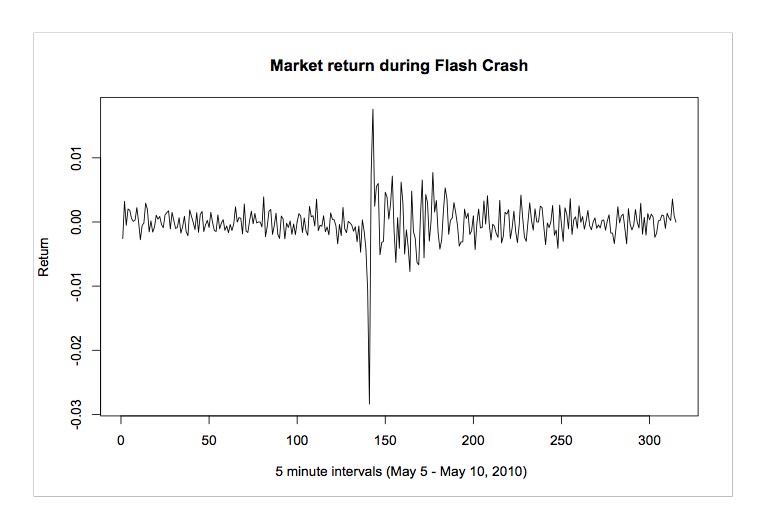} 
\caption{Intra day 5  minute returns for the time period around May 6 2010. Note how the large negative return triggers an increase in the volatility.}
\end{figure}

\section{ Intra day data: Cross-sectional properties }

We showed previously \cite{borland} that market panic on daily time scales is accompanied by high volatility, high dispersion, and low kurtosis. In addition the sum of signs S follows a bimodal distribution as a result of high correlations among stocks. Since it is difficult to calibrate and check our model very carefully  due to the lack of data and the relatively few instances of market panic historically, we instead look toward smaller time scales where we have many more observations. Our goal is to calculate the same quantities on these time scales and to see if the signatures of panic persist.

 A universe of  100 US stocks  (belonging to the SP100) was sampled at 15 minute intervals spanning the time period 2007-2009.  The market-wide average of this data is  shown in Figure 1.
Using this data, we calculated cross-sectional dispersion and cross-sectional kurtosis on 15 minute time scales. First, it is interesting to  notice the intra-day seasonality that these quantities exhibit. In particular, the average intra day dispersion is markedly bigger at the beginning of the day, decreasing as time goes on. This result was shown also in \cite{emerich} and more recently in \cite{jpintraday}. Kurtosis exhibits a less defined pattern, and the sum of signs $S$ shows some slight seasonality as well. These latter  observations are however not as marked as for the dispersion
(see Figure 2). More interesting details on the  relationship of dispersion, kurtosis and the index return are discussed in \cite{jpintraday}.

Bearing in mind the intra day seasonality, we show in Figure 3 the cross-sectional dispersion calculated at 15 minute time intervals for the entire time period considered. We also show cross-sectional kurtosis. Note that the correlation between the two does not exhibit the strong anti-correlation that we saw on daily scales (Figure 4). On the contrary, the correlation is positive at $30\%$.  Overnight data is omitted in this study, so one hypothesis is that the largest contribution to the decrease in kurtosis would come from overnight price moves. We discuss this point in more detail later on; in fact understanding why kurtosis and dispersion are negatively correlated on some time scales while not on others is an interesting feature which we try to understand and model \footnote{We also want to point out that our study of the cross-sectional kurtosis on daily time-scales  \cite{borland} was performed on a universe of 2000 US stocks, whereas the intra-day studies were just using 100 stocks.}

The next quantity that we look at is the normalized  sum of signs of returns  ($S$) over a given 15 minute interval.
We argued in \cite{borland} that $|S|$ is a good proxy for the correlations in the system. Intuitively, if $|S|$ is large (close to 1) , then all stocks will either go up or down together. Conversely, if $|S|$ is small,  there should be no particular co-movement among stocks. As a result, we expect to see  a uni-modal distribution of $S$ in normal times, and a bi-modal one in times of panic. This is what we observed on daily time scales. We studied the distribution of $S$ calculated from  our intra-day data and found the same behavior:  In times of panic, for example October 2008 (Figure 10), the distribution of $S$ is bimodal. In normal times, for example September 2008, the distribution of $S$ is unimodal (Figure 11).
The fact that this property appears on multiple time scales prompts the question: Is there memory in correlation?  This question was actually discussed in a recent paper \cite{multifrac} , but we argue here that their results  are not based on a fair definition of correlation. 

We shall expand on this in the next section, but before doing so, we wish to point out a possible criticism of our analysis: Note that \cite{JPcritic} a bimodal distribution can easily be obtained as an artifact  of conditioning on high volatility time points.  In such a case one "cherry-picks"  days (intervals) on which volatility was high, which by necessity corresponds to those days on which more stocks moved in the same direction, thus contributing to the higher absolute return over those particular days. We believe that this is not the case in our study; we do not"cherry-pick" certain days out of the sample, but are looking at the general behavior of  stock returns for all days in a region of interest.

\begin{figure}[t]
\label{fig12}
\includegraphics[width=4.5in]{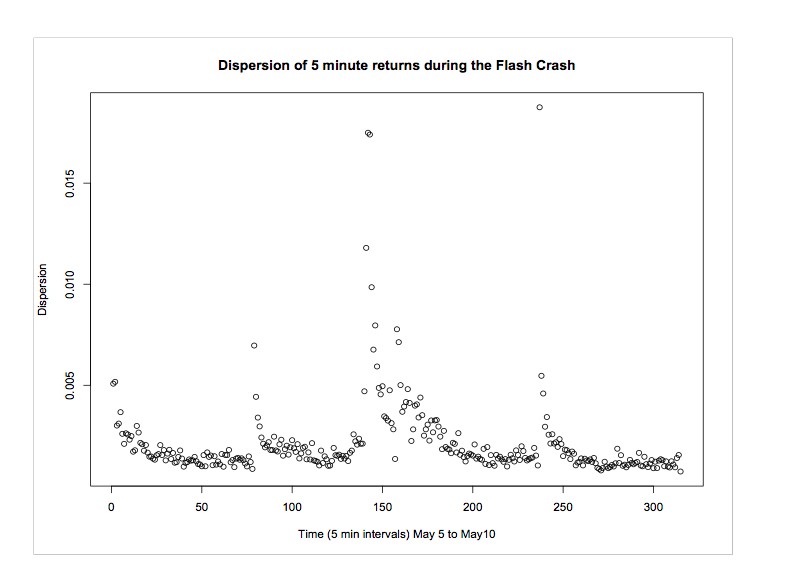} 
\caption{Intra day dispersion for the time period around May 6 2010, sampled at 5 minute intervals. Seasonality effects are apparent but nevertheless there is a jump in dispersion at the onset of  the Flash Crash}
\end{figure}

\begin{figure}[t]
\label{fig13}
\includegraphics[width=4.5in]{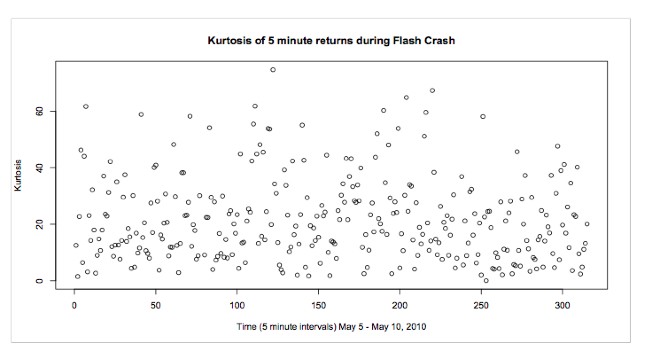} 
\caption{Intra day cross-sectional kurtosis  for the time period around May 6 2010, sampled at 5 minute intervals.  Nothing special happens at the onset of the Flash Crash. }
\end{figure}

\begin{figure}[t]
\label{fig14}
\includegraphics[width=4.5in]{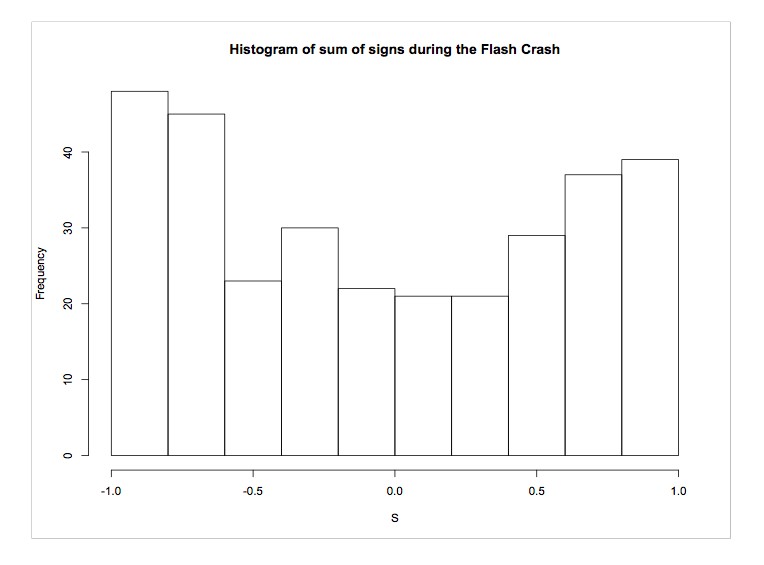} 
\caption{A histogram of the cross-sectional average sum of signs has a bi-modal distribution after the Flash Crash, indicating that even on this 5 minute time scale, correlations have increased and there are signs of panic. }
\end{figure}

\begin{figure}[t]
\label{fig15}
\includegraphics[width=4.5in]{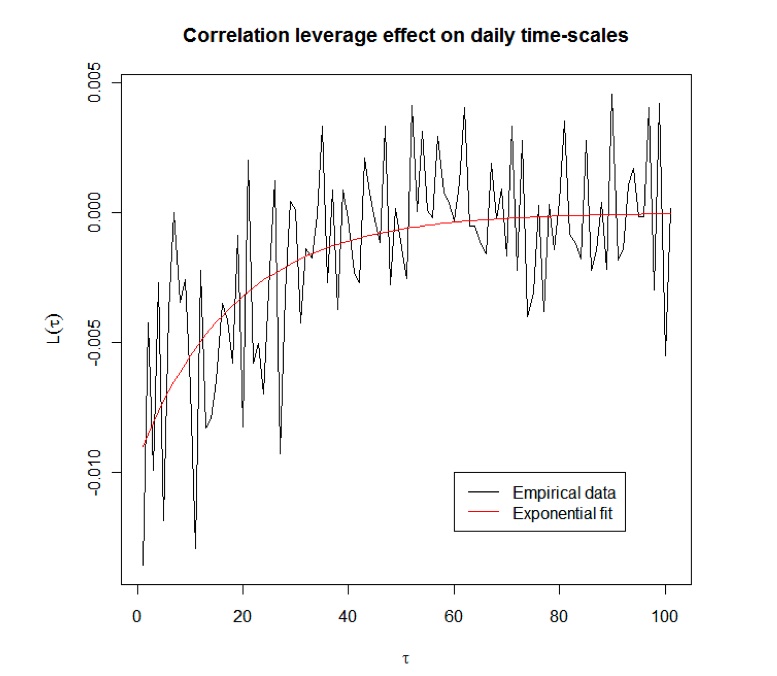} 
\caption{The correlation-leverage of Eq(\ref{eq:levaic}) for daily time-scales., well-fit by $L(\tau) = -A \exp{-\tau/T}$ with $A = 0.01 $ and $T = 18.4$. Negative returns imply higher correlations.  A similar result is found on intra-day time scales.}
\end{figure}

\begin{figure}[t]
\label{fig16}
\includegraphics[width=4.5in]{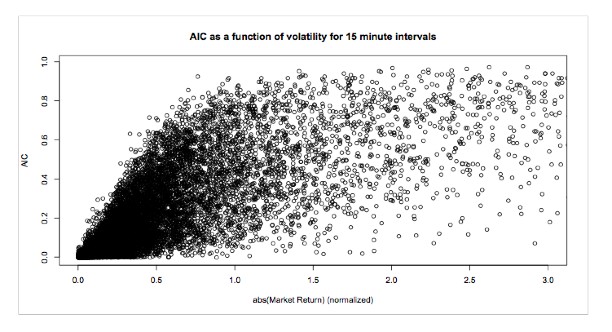} 
\caption{AIC as a function of market volatility. The scale is in terms of standard deviations. Notice that for volatilities greater than roughly 1 standard deviation, correlations are consistently higher than for volatilies less than 1 standard deviation. }
\end{figure}

\begin{figure}[t]
\label{fig17}
\includegraphics[width=4.5in]{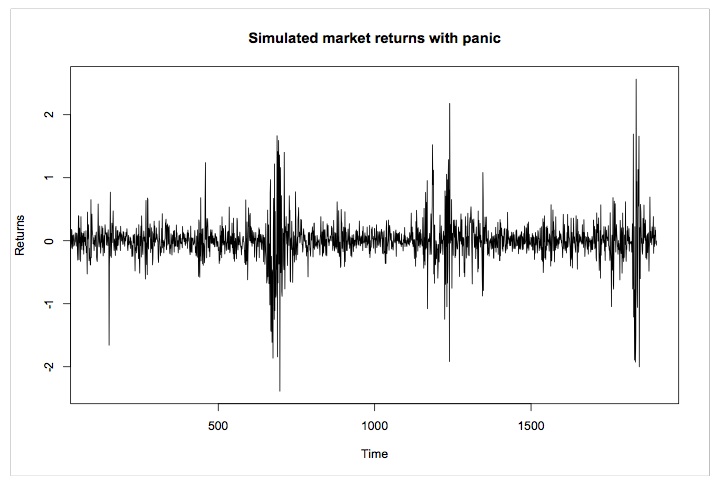} 
\caption{Market returns with periods of panic. At time 700, an endogenous shock is induced with an extreme negative return. At time 1150,  an exogenous volatility shock is applied. At around time 1800,
a spontaneous endogenous shock appears.}
\end{figure}

\begin{figure}[t]
\label{fig18}
\includegraphics[width=4.5in]{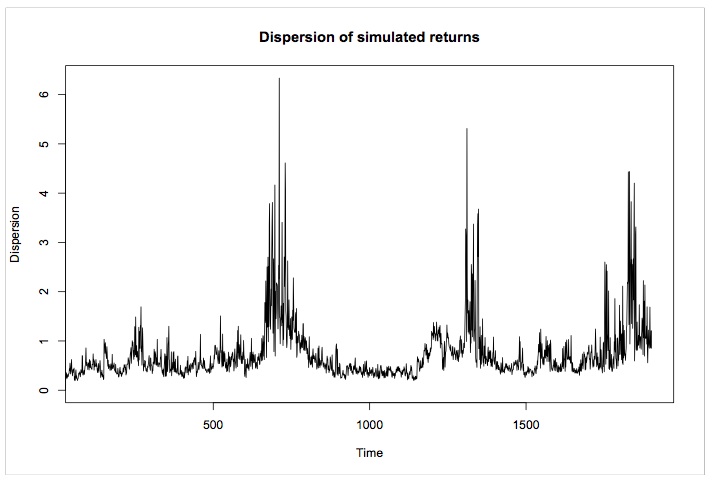} 
\caption{Dispersion increases during all three periods of panic.}
\end{figure}

\section{Memory  in correlations}

In a recent paper, \cite{multifrac}  studied the multi-fractal behavior of an average cross-market correlation defined from the sum of pair-wise correlations market-wide,
\begin{equation}
\label{eq:aicold}
AIC_{old} = \frac {2}{N(N-1)}\sum_i^{N-1}\sum_{j>i}^N \frac{(r_i-<r_i>)(r_j-<r_j>)}{\sigma_i \sigma_j}
\end{equation}

While interesting, upon closer inspection it is very clear that this measure is essentially the index volatility if the $\sigma_i$ and $\sigma_j$ are taken to be time series volatilities (see Figure 5). Hence, the nice results obtained for correlation dynamics are already
well-known from index volatility studies.

Instead we define a correlation measure in the spirit of \cite{multifrac}, yet which  we believe isolates correlation effects from volatility  effects. It is identical to Eq(\ref{eq:aicold}) ( assuming zero means), except that we take 
$\sigma_i = |r_i|$ and $\sigma_j = |r_j|$ resulting in  a sign-correlation function
\begin{equation}
\label{eq:aic}
AIC  = \frac {2}{N(N-1)}\sum_i^{N-1}\sum_{j>i}^N {(s_i)(s_j)}
\end{equation}

Note that any high frequency measure of correlation will suffer from the Epps effect \cite{epps}, which underestimates correlations due to asynchronous trading. However, we found that this could be safely ignored at the 15 minute time scale resolution.

Using the data set  of 100 US stocks described above,
we calculated at each time point the AIC as defined in Eq (\ref{eq:aic}). A plot of AIC is shown in Figure 6.  
 It is interesting to note that this quantity, 
for  a large number N  of stocks , is identically related to the square of a naive measure of correlations S which we defined in a  previous work,
 \begin{equation}
 S = \frac{\sum_i^N{sign(r_i)}}{N}
 \end{equation}
 Because of this correspondence, we shall sometimes speak of S, sometimes of AIC, but essentially both are  a proxy for the first-order correlation among stock returns.

Using this simple statistic as a quantifier for correlations, we  turn our attention to  the first of our questions, namely if there is memory in correlations. We  calculate the following quantity 
\begin{equation}
 V_{aic}(\tau)  = <(AIC(t)-AIC(t+\tau))^2>
\end{equation}
which is a variogram for the average  instantaneous correlation.
If there is no memory in the system, then  a plot of $V_{aic}$ as a function of the time lag $\tau$ will fluctuate around a straight line. However, this is not the case, and our results indicate that there is a power law decay of the instantaneous value of $AIC$ as time increases. Similar results for the memory in the volatility of individual stocks, as well as that of the volatility in the market index have already been presented in the literature \cite{jpandpotters}. Those quantities exhibit a long range memory with power law decays of index 0.2 and 0.34 respectively. In the case of AIC however, the memory is shorter-ranged , decaying with an index $\gamma$  empirically found to be  on average around 0.03.   Results for the memory on daily time scales are shown in Figure 7 ($\gamma = 0.05$).  In Figure 8,  the variogram of AIC sampled at 15 minute intervals is shown ($\gamma = 0.012$). The  evidence of memory  in the system is very clear, although the latter  results  contain a periodic pattern which is the effect of a convolution with  intra-day seasonality.
In addition to the variogram analysis,  we also calculated the multi-fractal spectrum for AIC, from which  there is evidence of  memory on different time-scales. 
Fianlly, we looked at the distribution of the sum of signs $S = sqrt{AIC}$ for 15 minute intervals, for the months of September 2008 and October 2008 (Figures 9 and 10). Within our framework,   the bimodal distribution in October  speaks to a panic phase.

\section{ The Flash-Crash}
On May 6 2010, market participants went in to a mini- panic as stocks suddenly exhibited drastic negative returns, only to recover fully by the end of the day. Much speculation to the cause of this behavior points a finger at high frequency electronic trading algorithms. But whatever the reason, it is clear that May 6 can be classified as a day of market panic, referred to as the Flash Crash. For us, this is an excellent scenario to study in the current framework.

In the days preceding and immediately following the Flash Crash, we sampled data for 100 highly capitalized US stocks at 5 minute intervals. We calculated cross-sectional dispersion and kurtosis, as well as $S$ and $AIC$. The plots in Figures 11-13   tell a nice story. First we show the 5 minute market returns on May 6 2010. One clearly sees how the volatility  jumps after the large negative return.
Simultaneously , dispersion (whose intra day seasonality is very clear) jumps to a high at the onset of the Flash Crash. Interestingly, kurtosis remains roughy the same. Regarding $S$, it is apparent that on and in the days following May 6 a bimodal distribution  is attained. This can be seen in Figure 14 where
a histogram of the cross-sectional average sum of signs after the Flash Crash is plotted, indicating that even on this 5 minute time scale, correlations have increased and there are signs of panic. Interpreting these results within the context of our model, in particular the fact that kurtosis doesn't decrease markedly in the Flash Crash, is the topic of the following section and was  also discussed in \cite{laurie}.

\section{ Endogenous or  Exogenous  Volatility Shocks: Negative Returns or External Fear?}

In our initial model \cite{borland} we hypothesized that market panic is triggered by an external volatility shock. In other words, a global "fear factor" pushes the financial system into a collective mode of highly correlated behavior, manifested by the fact that all stock returns will move in a highly similar manner, going either all up or all down together. This behavior is accompanied by an increase in volatility and an increase in cross-sectional dispersion. On daily time scales, it was also accompanied by a dramatic decrease in kurtosis. However, on intra day time scales we have seen that the cross-sectional kurtosis need in fact not decrease, while the other signatures of panic seem to  hold true also on these scales. 
We want to explore this phenomenon and understand it in a  little more detail.

Another point which we would like to understand better is if there is any asymmetry in the correlation dynamics,. Basically we want to see if there is evidence that  negative returns are more likely to trigger increases in correlation than positive returns. In fact, as we explore both of these points, we find a model that points to a distinction in the behavior of the cross-sectional  kurtosis depending upon how correlated   individual volatility shocks to  each stock are.   In other words, we shall see that the behavior is slightly different depending upon whether  the volatility shock is of a global (exogenous) nature, or whether it is 
more stock-specific or endogenous (for example  due to a large negative return to one or a few stocks in the market).

We now study the first of these questions, namely if there is asymmetry  in the correlation dynamics.
To this end we calculate the following quantity 
\begin{equation}
\label{eq:levaic}
L(\tau) = < AIC(t+\tau) r(t)>
\end{equation}
which we shall refer to as a correlation-leverage quantity. If   negative returns precede  an increase  in the  correlation AIC more frequently than  do positive returns, then this quantity will be negative. Indeed, a plot  of the correlation-leverage according to Eq(\ref{eq:levaic}) is shown in Figure 15, confirming that negative returns tend to precede increases in AIC. 
 In fact, this result is not surprising. Bouchaud et al already showed \cite{jpleverage} that there is s leverage effect present for returns in that negative returns precede increases in volatility. Also, \cite{fearfactor} discusses a similar result.
 
 Within the framework of our understanding in the current context,
it therefore makes sense that if negative returns imply higher volatility, then that should imply higher correlations because volatility and  the correlation dynamics are intricately related. This is for example illustrated in  Figure 16 where AIC is plotted as a function of market volatility (defined as the absolute value of the market return on a daily time scale). Notice that, as volatility increases, AIC grows until it saturates around 1. It is quite obvious from this plot that, on average, low volatility corresponds to low correlations, and high volatility corresponds to higher correlations. This plot is also consistent with the phase transition model of correlations which we proposed in \cite{borland}: For volatilities below a critical threshold, correlations are close to zero, with noise. This corresponds to the linear behavior at small volatilities. After a critical volatility, the  correlations take on high values, hence the flatter region at higher volatilities. 

Note that this interpretation is a contemporary one, and also could be turned on it's head: One could say that if AIC is large then the market volatility will be larger since the magnitude of  the sum of returns across stocks will likely be larger if all stocks go up or down in the same direction.  Nevertheless, if fits nicely with our phenomenological description of what is going on in the system.

\section{ A model: skew, volatility feedback, and correlation dynamics}
The stock returns for each instrument across time is modeled by \cite{jp&me}
\begin{equation}
\label{eq:one}
dy_t^i = \sigma_t^i d \omega_t^i
\end{equation}
where $y^i$ is the log stock price of the $i$-th stock, $\omega^i$ is a zero mean Gaussian noise with unit variance, and
\begin{equation}
\label{eq:two}
(\sigma_t^i)^2 = \sigma_0^2 [1 + \sum_{\tau = 1}^{\infty} \frac{g_{\tau}}{\sigma^2\tau} ( y^i_t - y^i_{t=\tau})^2 + \kappa \sum_{\tau = 1}^{\infty} \frac{g_{\tau}}{\sigma \sqrt{\tau}}(y^i_t - y^i_{t-\tau}) ]
\end{equation}
with $g_{\tau} = g/\tau^{\alpha}$. The parameters of the model are $g$, $\kappa$ and $\alpha$. Conditions for stability plus values which calibrate to real returns are discussed extensively in \cite{jp&me}. The feedback in the model is controlled by $g$, the power law memory in time is related to $\alpha$, and the skew and leverage effect relating negative returns and volatility depends on $\kappa$.
This model is motivated by the statistical feedback model that we presented in \cite{borland1,borland2,borland3} and is very similar to the Figarch models \cite{figarch,zumbach,Mandelbrot}.

For an ensemble of $N$ stocks, we assume that the $\omega^i, i = 1, \cdots, N$ are correlated proportional to $AIC= s^2$, where $S$ is the average sum of signs cross-sectionally. The empirical data that we showed in Figure 16 tells us that  
\begin{equation}
|S| = f(\sigma_M) + F_t
\end{equation}
where $\sigma_M$ is  the market volatility. One possible phenomenological manifestation of this relationship is the phase transition model which we proposed in \cite{borland}
\begin{equation}
\label{eq:s}
\frac{ds}{dt} = -a s -b  s^3 + F_t
\end{equation}
with $ a = \alpha(\sigma_c-\sigma_M)$,  $b$ a scaling parameter,  and $\sigma_c $ a critical volatility level. In this model, $S$ can be thought of as jiggling around in a potential well. 
When $\sigma_M < \sigma_c$,  that well only has one minimum at $0$, so $|S|$ fluctuates around that point.  If $\sigma_M > \sigma_c$, the potential well attains two new minima at $\pm \sqrt(a)$. $|S|  $becomes non-zero and correlations are high: Stocks tend to move up or down together in accordance, which is manifested in a bimodal distribution of $S$.We say that there is a phase transition as $\sigma_M$ crosses above the critical value, since the collective behavior of the stocks is qualitatively very different. This type of phase-transition model is based on the dynamics and theories  of self-organizing systems  as discussed in \cite{haken}.

The market volatility $\sigma_M$ can increase to values larger than $\sigma_c$ due to either  i) exogenous jumps (news) affecting all stocks so that $\sigma_0 $ becomes $\sigma_0 + \sigma_{shock}$, or ii) endogenous, idiosyncratic jumps which are more stock-specific and can for example be induced by a large negative return.

Finally we would like to point out that our model is for the scenario where the universe of traded instruments consists of stocks, mainly. If one for example includes ETFs  in a significant volume to the universe, correlations can be high even without panic but then one would expect that dispersion is lower \cite{Aevnine}. In that case, the stock dynamics could be written as 
\begin{equation}
dy_i = d(ETF) + \sigma_i(t) d\omega_i,
\end{equation}
where the correlations among the $\omega_i $ still may follow Eq(\ref{eq:s}).  But because of the ETF exposure, there would be a necessarily high correlation among all stock returns if the volume of traded ETF is high, which is actually the present case with about 9$\%$ market volume in this instrument.

\begin{figure}[t]
\label{fig19}
\includegraphics[width=4.5in]{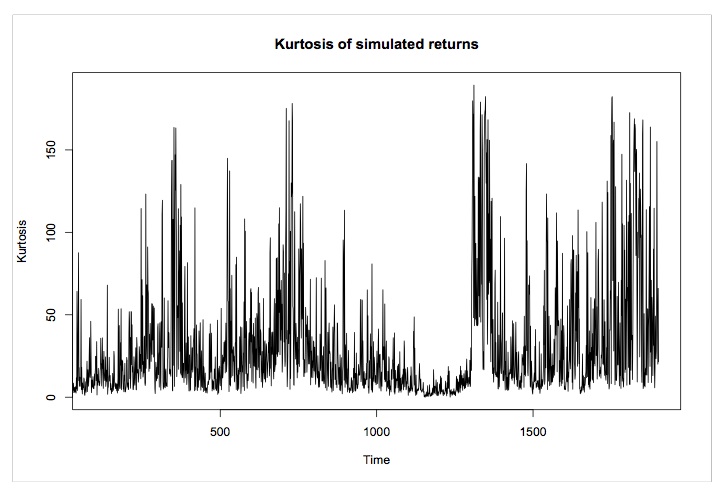} 
\caption{Kurtosis  decreases drastically only when the panic was induced by an exogenous volatility shock (around time 1150).}
\end{figure}

\begin{figure}[t]
\label{fig20}
\includegraphics[width=4.5in]{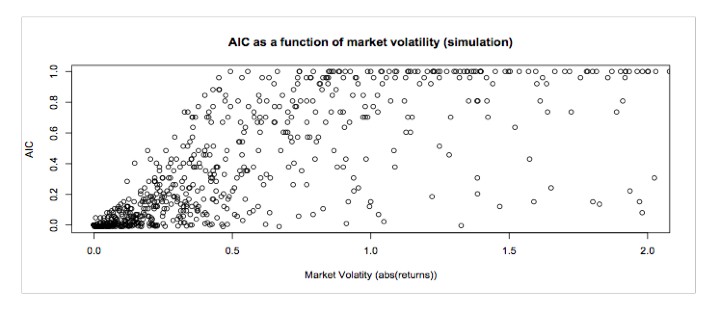} 
\caption{A plot of AIC versus market  volatility.}
\end{figure}

\begin{figure}[t]
\label{fig21}
\includegraphics[width=3.5in]{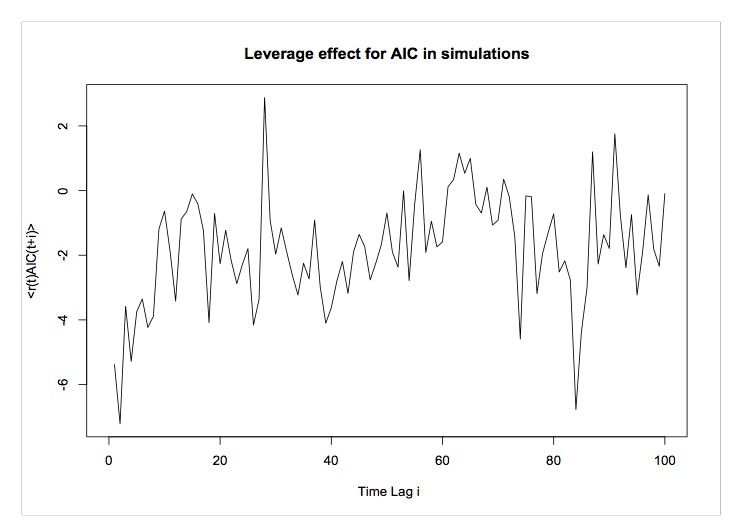} 
\caption{The correlation leverage effect in our simulations.}
\end{figure}

\section{Simulations}

To illustrate our model in a qualitative fashion we ran a  simulation using Eq(\ref{eq:one}) and Eq(\ref{eq:two}) as the basis for the dynamics of each stock. Here, the feedback parameter $g=0.35$, the skew parameter $\kappa = 0.15$ and $\alpha = 1.15$. The base volatility was chosen at $\sigma_0 = 0.20$, and we included only up until $N=100$ terms in the volatility feedback term. Our goal was not to calibrate to real data here, but rather to attain a process which qualitatively captures the main stylized facts that real stock returns exhibit over time. For the cross sectional dynamics we modeled the correlations across stocks at a given time point with the phenomenological equation Eq(\ref{eq:s})
using $b = 0.01$ and  $F_t  = 0.1\tilde{\nu}$  where $\tilde{\nu}$ is a standard Brownian noise. The critical volatility $\sigma_c$ was taken to equal one standard deviation of the historical market volatility, itself measured as the squared variation of historical returns over the past 100 time points.  At each time point in the simulation, an estimate of the current market volatility was obtained  as the absolute value of the recent market return.

In this setting, the current volatility can change either due to a "global" shock such that $\sigma_0$ effectively becomes $\sigma_0 + \sigma_{shock}$, or due to a large negative return in one or more stocks. The first case corresponds to a market wide exogenous fear factor, such as happened around the Lehman collapse.  The second case better describes an endogenous, event driven panic such as happened in the Flash Crash.

In our simulations we can induce both types of volatility shocks, and look to see the effect on the resulting cross-sectional properties which we have discussed in the previous sections.  Figure 17 shows the simulated market returns, with three periods of panic (one induced by a negative return  of  4 standard deviations magnitude  at time 700, one induced by a volatility shock of size $\sigma_{shock} = 0.4$ at time 1150, and one which occurred spontaneously toward the end). 

The first  period of panic  was induced by a large endogenous  negative return in one stock, which propagated up to the market level. The dispersion increased as exhibited in Figure 18, while the kurtosis (Figure 19)  did not decrease but was instead somewhat positively correlated with the dispersion. This case mimics the scenario of the Flash Crash. On the other hand, as also shown in the  plots of Figure 18 and Figure 19, the dispersion and kurtosis of the  region where the panic was induced by a global exogenous shock exhibit a strong negative correlation, as in for example the crash of October 2008. 
In addition, the dynamics spontaneously induced another period of panic which follows the statistics of an endogenous shock. In all cases, histograms of S are bimodal during the times of panic, and unimodal in normal times, as already shown in \cite{borland}. Furthermore, a plot of $AIC$ versus volatility (Figure 20)  is qualitatively similar to our empirical observations. And finally, the simulated data exhibits  a  correlation-leverage effect  much as that observed empirically (see Figure 21).

\section{Conclusions}
 We have explored the dynamics of a simple correlation statistic, namely the sum of signs of cross-sectional return, or equivalently, its square which is closely related to the cross-sectional sign correlation. We have shown that this quantity exhibits a memory over time  reinforcing the idea that correlations tend to cluster, although by far not as notably as volatility. We also showed that there is a kind of leverage-correlation effect (akin to the leverage effect between volatility and negative returns), in that negative returns precede an increase in the correlations.  In a previous paper we had studied statistical signatures of market panic on daily time scales, and had found that in such instances, market volatility rises, as does cross-sectional dispersion and correlation among stocks, while the cross-sectional kurtosis exhibited a marked decrease. We proposed that the drastic change in the cross-sectional dynamics of markets in a panic situation can be modeled as a phase transition, induced by an external volatility shock. 
 
  Here we  extended this study to intra-day time scales and found that all signatures of panic remain the same, except that the kurtosis does not necessarily decrease during a crash. This was exemplified by an analysis of the cross-sectional dynamics on and around May 6, 2010,  when markets crashed only to rebound fully shortly after, hence coined the Flash Crash. 
   The subtle difference in the cross-sectional distribution of stock returns during this crash as opposed to for example the panic of October 2008 can be attributed to the source of the volatility shock which induces  the crash. In particular we found that if the shock is the result of an endogenous, idiosyncratic behavior, such as an unusually large negative return in one or more stocks, then market volatility, dispersion and correlations will increase, but kurtosis  doesn't necessarily decrease. This can be understood  as follows. If the volatility shock is global or market wide, as would be the case from an exogenous news event, then the individual volatilities of each stock will become more similar, which leads cross-sectionally to a more Gaussian distribution, and low kurtosis.. On the other hand, if the volatility shock affects just a few stocks, then the cross-sectional distribution reflects that of an ensemble of random variables each with a different volatility, yielding a high kurtosis.  
     Interestingly we also found that in general kurtosis and dispersion are not strongly anti-correlated on intra-day timescales, even for periods such as late 2008 where they exhibit a strong negative correlation on daily timescales. This difference would be attributed to overnight price moves which, in the framework of our model,  would impound  globally correlated, exogenous volatility into the system.  
     
   We proposed a model based on \cite{borland} that includes skew in the dynamics  for each individual stock and were able to reproduce in simulations the basic dynamics and statistical properties of these two different sources of panic, including the leverage-correlation effect which we found.  The work presented here is entirely qualitative in nature, and a future study should attempt to calibrate the model to empirical data. This could be useful to increase our understanding, but also for predictive purposes which could have many applications for risk control and the management of extreme correlation events.

\end{document}